\documentclass[fleqn,twoside]{article}
\usepackage{espcrc2}

\usepackage{graphicx}
\usepackage[figuresright]{}

\newcommand{\AmS}{{\protect\the\textfont2
  A\kern-.1667em\lower.5ex\hbox{M}\kern-.125emS}}

\title{Steps towards a map of the nearby universe}

\author{R. D'Abrusco \address{Dept. Physical Sciences, University 'Federico II',
              Naples, Italy \& Institute of Astronomy, University of Cambridge}, 
              G. Longo, A. Staiano, E. de Filippis, M. Paolillo \address{Dept. Physical Sciences, University 'Federico II',
              Naples, Italy \& INAF-Italian National Institute of Astrophysics, via del Parco
              Mellini, Rome, I  \& INFN -
              Napoli Unit, Dept. of Physical Sciences, via Cinthia 9, 80126,
              Napoli, ITALY}, \& M. Brescia \address{INAF - Osservatorio Astronomico di Capodimonte, via Moiariello 16, 80131, Napoli, Italy \& INFN -
              Napoli Unit, Dept. of Physical Sciences, via Cinthia 9, 80126,
              Napoli, ITALY}}          
\begin{document}

\begin{abstract}
\vspace{1pc}
We present a new analysis of the Sloan Digital Sky Survey data 
aimed at producing a detailed map of the nearby ($z<0.5$) universe. 
Using neural networks trained on the available spectroscopic base of 
knowledge we derived distance estimates for $\sim 30$ million galaxies
distributed over $\sim 8,000$ sq. deg. 
We also used unsupervised clustering tools developed in the framework 
of the VO-Tech project, to investigate the possibility to understand 
the nature of each object present in the field and, in particular, to 
produce a list of candidate AGNs and QSOs. 
\end{abstract}
\maketitle

\section{Introduction}
The recent developments in the fields of wide field digital 
detectors, dedicated survey telescopes and computer sciences, 
promise to fulfill in a few years one of the oldest dreams of the 
scientific community, namely, the production of an accurate 
taxonomy of the nearby universe. 
At a very basic level, such taxonomy will consist in a detailed 
description of both positions and types for all objects matching 
well defined selection criteria (e.g. flux limited or volume 
limited samples).
Even so, however, it will be of the uttermost relevance for 
many fields of cosmology and astroparticle physics 
\cite{cuoco_2006,astronomy}.  
In what follows we shall shortly outline the first results of our 
ongoing efforts to produce a 3-D map of the nearby universe with a 
characterization of galaxy types in a few, broadly defined, 
categories: normal (early and late type) galaxies, AGN, QSO, 
etc.
Such work takes place in the framework of the European VO-Technological 
Infrastructures project (VO-Tech, \cite{votech}).
We used as input data the Sloan Digital Sky Survey Data Release 4
and/or 5 (hereafter SDSS4/5; \cite{SDSS5}) which provides photometric 
data in 5 bands for several hundred million galaxies distributed 
over $\sim 8,000$ sq. deg. and additional spectroscopic 
information for a subsample of about 1 million extragalactic objects. 
\section{The photometric redshift}\label{redshifts}
In order to evaluate photometric redshifts we made use of an 
improved version \cite{dabrusco_2006} of the Neural Networks 
(NNs) method presented in \cite{tagliaferri_2003,firth_2003}. 
Both steps were accomplished using the NN architecture known 
as MLP (Multi Layer Perceptron, \cite{bishop_1995,duda}).

We adopted a two steps approach: first we trained an MLP to 
recognize nearby (id est with redshift $z<0.25$) and distant 
($z>0.25$) objects, then we trained two separate MLPs to work 
in the two different redshift regimes. 
Such approach finds a strong support in the fact that in the 
SDSS-5 catalogue, the distribution of galaxies inside the two 
different redshift intervals is dominated by two different galaxy 
populations: the Main Galaxy (MG) sample in the nearby region, 
and the Luminous Red Galaxies (LRG) in the distant one. 
The use of two separate networks ensures that the NNs achieve 
a good generalization capabilities in the nearby sample, leaving 
the biases mainly in the distant one.
To perform the separation between MG and LRG objects, we extracted 
from the SDSS-4 Spectroscopic Subsample (hereafter, SpS) training, 
validation and test sets weighting, respectively, $60\%$, $20\%$ 
and $20\%$ of the total number of objects (449,370 galaxies).
The resulting test set, therefore, consisted of 89,874  randomly 
extracted objects. 
After this first step, the evaluation of photometric redshifts 
was performed working separately in the two regimes.
Errors were then evaluated on the test set by measuring 
the dispersion of points in the $z_{spec}$ vs $ z_{phot}$ plane, 
i. e. the variance of the $z_{spec}-z_{phot}$ variable, 
after performing an interpolative correction to correct for 
residual systematics. 

\begin{figure}
\centering
\includegraphics[width=7.0cm]{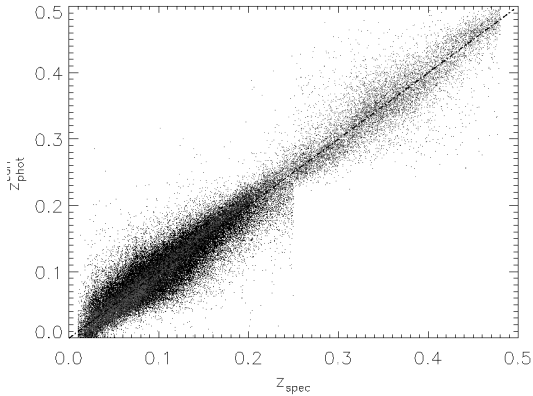}
\caption{Distribution of spectroscopic versus photometric redshifts in the test set. 
Lighter grey points mark LRG galaxies. Notice the larger scatter of non-LRGs in the distant 
($z>0.25$) sample.}
\label{fig:redshifts}
\end{figure}

Our results can be summarized as follows:
\begin{itemize}
\item For the MG sample, the robust variance turned out to be 
$\sigma_3 = 0.0208$ over the whole redshift range and $0.0197$ and
$0.0245$ for the nearby and distant objects, respectively.
\item For LRG sample we obtained $\sigma_3 \simeq 0.0163$ over the whole range,
and $\sigma _3 \simeq 0.0154$ and $\sigma _3 \simeq 0.0189$ for the nearby and distant samples,
respectively.
\end{itemize}
\section{The clustering}\label{clustering}
The implementation of a reliable classification scheme requires the 
partitioning of the observed parameter space in clusters of objects 
sharing some underlying common physical property.
Obviously, since there is no a priori reason to assume that photometric 
classifications must reflect strictly any morphological classification, 
any effective classification method must be unsupervised and 
partition the photometric parameter space using only the statistical 
properties of the data themselves.

Most unsupervised methods, however, require the number of clusters to be
provided \emph{a priori}. 
This circumstance represents a serious problem when exploring large and 
complex data sets where the number of clusters can be very high or, in 
any case, unpredictable. 
A simple treshold criterium is not satisfactory in most astronomical
applications due to the high degeneracy and the noisiness of the
data which lead to the erroneous agglomeration of data.
We therefore implemented a hierarchical approach which starts from a 
preliminary clustering performed using as unsupervised clustering algorithm 
the so called "Probabilistic Principal Surfaces"  or PPS described in
\cite{chang_2000,staiano_2003}, and then makes use of the Negative Entropy 
concept and of a dendrogram structure to agglomerate the clusters found in 
the first phase \cite{dabrusco_2006}. 

\begin{table}
\begin{center}
\label{tabella}
\begin{tabular}{@{} c| c c c c c c c }
\hline
 Cl. n    & SP0  & SP1& SP2& SP3& SP4& SP6 \\
\hline
1 &69 &145 &9362 &48 &0 &12\\
2 &25 &133 &13370 &10 &0 &12\\
3 &149 &132 &63 &64 &0 &5\\
4 &44 &3396 &1530 &189 &67 &1\\
5 &202 &85 &447 &2428 &6 &10\\
6 &26 &125 &13728 &12 &0 &12\\
7 &0 &0 &0 &0 &0 &484\\
8 &1 &1 &1 &0 &0 &329\\
9 &541 &1507 &127 &4750 &18 &1\\
10 &89 &474 &2117 &19 &4 &529\\
\hline
\end{tabular}
\end{center}

\caption{Distribution of a subsample of objects in the most significant 
clusters. Columns correspond to different values of the 
\emph{specClass} index provided by the SDSS.}
\end{table}

We first applied the PPS algorithm to the sample of 
spectroscopically selected SDSS DR-4 objects using as parameters 
for the clustering the 4 colours obtained from model magnitudes 
(u-g,g-r,r-i,i-z). 
We fixed the number of latent variables and latent bases of the
PPS to 614 and 51 respectively, so obtaining at the end of this
step 614 clusters, each formed by objects which only respond to a
certain latent variables. 
We choose a large number of latent variables in order to obtain 
an accurate separation of objects and to avoid that any group of 
distinct but near points in the parameter space could be projected 
onto the same cluster by chance.
These first order clusters were then fed to the NEC algorithm
which determined the final number of clusters. 
The plateau analysis of the agglomerative process and the inspection of the 
dendrogram allowed to set the treshold to a value corresponding to 31 clusters. 
We present in table (\ref{tabella}) the ten most populous clusters 
together with the distribution of the \emph{specClass} index within each 
cluster. 
The additional 21 clusters represent less than 10 $\%$ 
of the objects and are still under investigation.
It needs to be emphasized that the clustering makes use of the photometric 
data only and the spectroscopic information is used only to validate them.
As it can be seen, galaxies (SP2) clearly dominate
clusters 1, 2 and 6. 
Whether this separation reflects or not some deeper differences among the 
three groups (such as, for instance, different morphologies), cannot be 
assessed on the grounds of presently available data. 
AGNs (SP3) dominate clusters 5 and 9 even though some contamination from
galaxies  exists. 
Late type stars (SP6) populate mainly clusters 7 and 8 and are strong contaminants 
of cluster 10 which also is dominated by  galaxies. 
It needs to be stressed, however, that 
the use of the \emph{specClass} as label must be taken with some caution since 
it is becoming increasingly evident that this index suffers from 
severe biases \cite{rifatto_2006}.

\noindent {\it Acknowledgements:} this work was sponsored by the
Italian MIUR through a PRIN grant and by the EC through the VO-Tech 
project. 
The authors wish to thank G. Miele and G. Raiconi for 
useful discussions.

%References should be collected at the end of your paper. Do not begin
%them on a new page unless this is absolutely necessary. They should be
%prepared according to the sequential numeric system making sure that
%all material mentioned is generally available to the reader. Use
%\verb+\cite+ to refer to the entries in the bibliography so that your
%accumulated list corresponds to the citations made in the text body. 

%Above we have listed some references according to the
%sequential numeric system \cite{Scho70,Mazu84,Dimi75,Eato75}.

\end{document}